\documentclass[%
 reprint,
 superscriptaddress,
amsmath,amssymb,
]{revtex4-1}

\usepackage{graphicx}
\usepackage{dcolumn}
\usepackage{bm}
\usepackage{hyperref}
\usepackage[mathlines]{lineno}
\usepackage{multirow}
\usepackage{boldline}
\usepackage{xcolor}

\begin{document}

\title{Dynamic Vulnerability in 
Oscillatory Networks and Power Grids}

\author{Xiaozhu Zhang}
\email{xiaozhu.zhang@tu-dresden.de}
\affiliation{Chair for Network Dynamics, Institute for Theoretical Physics and Center for Advancing Electronics Dresden (cfaed), Technical University of Dresden, 01062 Dresden, Germany}
\author{Cheng Ma}
\affiliation{School of Physics, Nankai University, Tianjin 300071, China}
\author{Marc Timme}
\email{marc.timme@tu-dresden.de}
\affiliation{Chair for Network Dynamics, Institute for Theoretical Physics and Center for Advancing Electronics Dresden (cfaed), Technical University of Dresden, 01062 Dresden, Germany}

\date{\today}

\begin{abstract} 
Recent work found distributed resonances in driven oscillator networks and AC power grids. The emerging dynamic resonance patterns are highly heterogeneous and nontrivial, depending jointly on the driving frequency, the interaction topology of the network and the node or nodes driven. Identifying which nodes are most susceptible to dynamic driving and may thus make the system as a whole vulnerable to external input signals, however, remains a challenge. Here we propose an easy-to-compute Dynamic Vulnerability Index (DVI) for identifying those nodes that exhibit largest amplitude responses to dynamic driving signals with given power spectra and thus are most vulnerable. The DVI is based on linear response theory, as such generic, and enables robust predictions. It thus shows potential for a wide range of applications across dynamically driven networks, for instance for identifying the vulnerable nodes in power grids driven by fluctuating inputs from renewable energy sources and fluctuating power output to households.

\end{abstract}

\maketitle

\section{Introduction}

Oscillatory networks, modeling the underlying mechanisms of many real-world systems ranging from gene and neural circuits \cite{hussain2014,jahnke2014} to AC power grids \cite{kundur1994,filatrella2008,rohden2012,motter2013,dorfler2013,coletta2016}, exhibit highly nontrivial responses to external driving signals \cite{zanette2004,zanette2005,tyloo2018,zhang2019,haehne2019}, due to the complexity in the underlying topology and the nonlinearity in the coupling function. Recently, growing attention has been drawn to the topic of dynamically driven networks, in part because of the important application of the second-order Kuramoto-type oscillator model in power grid operation and control \cite{zhang2019,schaefer2017,kettemann2016,kettemann2018,haehne2019}. With an increasing share of fluctuating renewable energy sources integrated in modern power grids, it is crucial for grid operators to predict the distributed frequency responses to systematic and stochastic fluctuations, to identify which units are most susceptible and may thus make the system as a whole vulnerable to dynamic inputs.

For the example of power grid models, key aspects of network responses to dynamical perturbations have been uncovered recently: about the impact of various types of perturbation signals, including the scaling in the relaxation of power grids after pulse-like perturbations \cite{kettemann2016,kettemann2018}, the differential response to static perturbations\cite{manik2017network}, the distributed dynamic patterns in response to dynamic perturbations \cite{zhang2019}, the fluctuation-induced non-Gaussian grid frequency distribution \cite{schaefer2018} and the escape of a system from an operation state if driven by white noise \cite{schaefer2017} and/or non-Gaussian noises \cite{tyloo2019,hindes2019}. Specifically, the time-averaged nodal deviations from the network mean response was ranked using a centrality measure based on the Laplacian spectrum \cite{tyloo2018key}. For lossy networks, averaged nodal sensitivity to fluctuations across network have been numerically investigated \cite{auer2017} and estimated via the nodal variance \cite{plietzsch2019}. Yet, it is still unclear which stochastic signals may cause network-wide response patterns and how to quickly and precisely identify those nodes that potentially exhibit most severe responses and thus are most vulnerable to such perturbations.

The core of the puzzle lies in an intriguing phenomena of dynamic network resonance\cite{zhang2019}, which is present in oscillator models with two (or more) variables per node (such as the second order Kuramoto model, an oscillatory power grid model) but not in the networks of phase oscillators such as in the original Kuramoto model \cite{zanette2004,zanette2005,tyloo2018,zhang2019}. While the network responses for low- and high-frequency signals are trivial thus fairly predictable---homogeneous responses for slowly-changing signals and localized responses for fast-changing signals---fluctuations in the resonance frequency regime of a network system induce complex resonance patterns in oscillatory networks \cite{zhang2019}. The patterns are jointly determined by the perturbation frequency, the underlying network topology, the initial unperturbed network state (base operating state), and the location of the perturbation and the response of interest. Although the resonant responses can be deterministically and precisely computed for given perturbation time series deriving and evaluating a linear response theory \cite{zhang2019}, a straightforward, fast and reliable method for estimating the resonant response strengths for stochastic signals is still missing, mainly because such signals contain an extended band of frequencies.

Here, going beyond structural vulnerability in networks, we propose the Dynamic Vulnerability Index (DVI), a computationally inexpensive vulnerability measure to assess and to rank the largest possible resonant response of individual nodes in oscillatory networks. The networks are driven by stochastic perturbations containing a characteristic power spectral density (PSD) function. 
In power grid research and beyond, the term \textit{network vulnerability} is typically used to describe the impact of purely topological changes on network performance  \cite{arianos2009,simonsen2008,albert2004,song2005,qu2002}. The meaning of the \textit{vulnerability} of a node was extended to considering the node's transient response to a pulse-like perturbation \cite{gutierrez2011,tyloo2019vulnerability}, and recently to the time-averaged response to stochastic perturbations \cite{tyloo2018key}. Here we propose a Dynamic Vulnerability Index (DVI) that expands the definition by considering the global maximum of a node's \textit{dynamic} response to a \textit{stochastic} input signal.
Employing a linear response theory\cite{zhang2019} and a frequency-specific estimate of the resonant response strength, the DVI exhibits high prediction power and helps to identify those nodes potentially respond most strongly to a stochastic resonant perturbation and thereby posing systemic risks in power grid stability. Especially, the DVI identifies the vulnerable nodes at unexpected locations in the network not foreseeable from the topology alone.

\section{Resonant network response patterns}

Consider a network of $N$ second-order Kuramoto-type oscillators with dynamics governed by 
\begin{equation}
    \ddot{\theta}_i=P_i-\alpha\dot{\theta}_i+\sum_{j=1}^NK_{ij}\sin(\theta_j-\theta_i)+\delta_{ik}D(t) 
\end{equation}
and driven by an external fluctuating signal $\delta_{ik}D(t)$ only present at node $k$. Here $\theta_i$ and $P_i$ denote the rotation angle and the natural acceleration of oscillator $i$ (proportional to power input or output at $i$), $\alpha>0$ parametrizes the damping coefficient and $K_{ij}>0$ the coupling strength of node pair $(i,j)$. The model is equivalent to a coarse-grained model of AC power grids \cite{zhang2019} enabling effective inertia for units with fluctuating power input from renewables. It describes the collective dynamics of $N$ sub-grids with an effective damping coefficient $\alpha$ rotating at grid frequency $\Omega_0=2\pi\times50$ Hz (or $60$ Hz in the US and parts of Japan) in the normal operation state. In this context $\theta_i$ represents the center-of-inertia angle deviation of sub-grid $i$ to the reference frame rotating at $\Omega_0$, $P_i$ the power generated ($P_+>0$) or consumed ($P_-<0$) in sub-grid $i$ and $K_{ij}$ the line capacity of the power transmission between sub-grid $i$ and $j$. The driving signal $\delta_{ik}D(t)$ is a fluctuating time series additive to the average power generation or consumption at sub-grid $k$. For the purpose of the modelling setup each sub-grid is presented as one node of a graph.

\begin{figure}[h]
	\centering
	\includegraphics[width=\columnwidth]{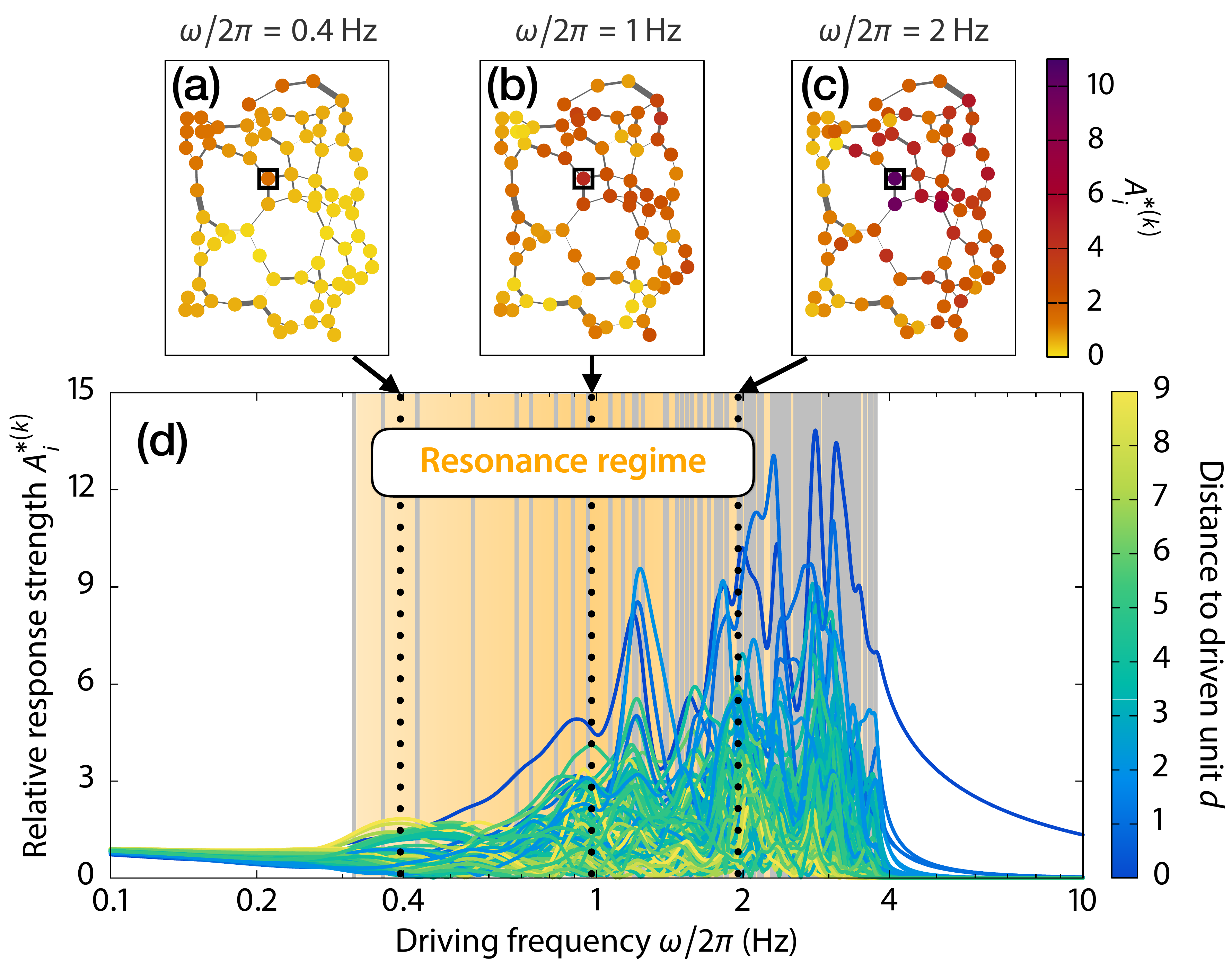}
	\caption{\textbf{Complex network resonance patterns and large amplitude responses.} (a$_1$-a$_3$) Distinct network resonance patterns for three driving frequencies. Each node is color-coded by the relative response strength $A^{*(k)}_i:=|\dot{\Theta}_i^{(k)}|/\lim_{\omega\rightarrow0}|\dot{\Theta}_i^{(k)}|$ and the driven node is marked with a black square. In simulation we use per-unit parameters $\alpha=\Omega_0\bar{D}/2\bar{H}$, $P_+=\Omega_0/2\bar{H}$, $P_-=-P_+/3$, $K_{ij}=2P_+$ with the normalized damping coefficient $\bar{D}=0.02\,\mathrm{s}^{2}$ and the aggregated inertia constant $\bar{H}=4\,\mathrm{s}$. The network topology is created by a random growth model of power grid networks \cite{schultz2014}.(b) The resonant response patterns varies with driving frequency in the resonance regime (shaded orange). The relative response strength of individual nodes are color-coded by its distance to the driven unit. Grey vertical lines indicate the $N-1$ resonance frequencies $\omega^{[\ell]}_{\text{res}}$. }
	\label{fig:resonance}
\end{figure}

How does such a network respond to dynamic input signals (Fig.~\ref{fig:resonance})? Close to a normal operation state of the power grid, i.e. a stable fixed point $\bm{\theta}^*:=\left(\theta_1^*,\cdots,\theta_N^*\right)$ of the oscillatory network, the collective response $\bm{\Theta}(t):=\bm{\theta}(t)-\bm{\theta}^*$ to a perturbation vector $\mathbf{D}^{(k)}(t)$ defined via its components $D^{(k)}_i(t):= \delta_{ik}D(t)$ is accurately given by a linear response theory \cite{zhang2019}
\begin{equation} \ddot{\mathbf{\Theta}}^{(k)}=-\alpha\dot{\mathbf{\Theta}}^{(k)}-\mathcal{L}\mathbf{\Theta}^{(k)}+\mathbf{D}^{(k)},
\end{equation}
where $\mathcal{L}$ with $\mathcal{L}_{ij}:= K_{ij}\cos(\theta_j^*-\theta_i^*)$ for $i\neq j$ and $\mathcal{L}_{ii}=-\sum_{j\neq i}\mathcal{L}_{ij}$ is a weighted graph Laplacian. The linear network response is analytically solvable by projecting it to the orthogonal eigenspaces of the Laplacian matrix \cite{zhang2019}. For a sinusoidal perturbation $D(t)=\varepsilon e^{i(\omega t+\varphi)}$, the frequency response at unit $i$ to a perturbation at unit $k$ reads
\begin{equation}
	\dot{\Theta}_i^{(k)}(\omega,t)=\varepsilon e^{\imath(\omega t+\varphi)}
	\sum_{\ell=0}^{N-1}\dfrac{i\omega v^{[\ell]}_kv_i^{[\ell]}}
	{-\omega^2+\imath\alpha\omega+\lambda^{[\ell]}}.
	\label{eq:ResponseSinusoidal}
\end{equation}
Here $\lambda^{[\ell]}$ and $v_i^{[\ell]}$ denote, respectively, the $\ell$-th eigenvalue and the $i$-th component of the corresponding eigenvector. The eigenvalues are indexed as $0=\lambda^{[0]}\leq\cdots\leq\lambda^{[N-1]}$. When one of the eigenmodes is excited, that is, when the perturbation frequency $\omega$ maximizes the contribution of an eigenmode in Eq.~\ref{eq:ResponseSinusoidal} with an eigenfrequency
\begin{align}
	\omega=\omega^{[\ell]}_{\text{res}}:=\sqrt{\lambda^{[\ell]}-\dfrac{\alpha^2}{4}},
	\label{eq:ResonanceFrequencies}
\end{align}
the network response is dominated by the resonant eigenmode characterized by an overlap factor $v^{[\ell]}_kv_i^{[\ell]}$, constituting a nontrivial, highly heterogeneous dynamic pattern (Fig.~\ref{fig:resonance}).

We emphasize that network resonances emerge not only at a single perturbation frequency, but at all frequencies in a wide frequency range we call the resonance regime, i.e. 
\begin{equation}
    \omega\in I_{\mathrm{res}}:=\left[\sqrt{\lambda^{[1]}-\frac{\alpha^2}{4}},\sqrt{\lambda^{[N-1]}-\frac{\alpha^2}{4}}\right].
\end{equation} 
Extraordinarily high strengths of frequency response up to an order-of-magnitude (e.g. 12 times) larger than the homogeneous response strength in the low frequency limit \cite{zhang2019} may appear across the network (Fig.~\ref{fig:resonance}d). Furthermore, the resonant response pattern sharply depends on the driving frequency. In an exemplary network, the response pattern appears to be distinctly different for three frequencies with no more than $1$Hz apart from each other (Fig.~\ref{fig:resonance}a-c). Beside the heterogeneity in response amplitude, each node's response additionally exhibits heterogeneous phase delay towards the perturbation signal, due to the characteristic arguments of the complex responses \eqref{eq:ResponseSinusoidal}.

\section{Indexing resonant responses}

Even perturbed only by a single-frequency resonant signal, the network already exhibits complex response patterns in terms of the strength and the phase delay of the sinusoidal response \eqref{eq:ResponseSinusoidal}. In reality, power grids are constantly exposed to noisy fluctuations in  renewable power generation and in the power consumption of households and industry, which consist of Fourier components with a wide range of frequencies in the resonance regime, stochastic magnitudes and random phases. For a given noisy perturbation time series, the network response time series is computable by summing up the linear response to each frequency \eqref{eq:ResponseSinusoidal}. However, the influence of future remains unknown. Making general predictions for a network's complex resonant response to noisy fluctuations, and particularly, identifying the most susceptible nodes still remains a challenge.

We propose an index of the vulnerability of individual nodes in a network under resonant perturbations, the Dynamic Vulnerability Index (DVI), which helps to rank the maximum resonant response magnitude for perturbations with a characteristic PSD $S(\omega)$. Measurements of wind and solar power systems indicate that both of the strongly fluctuating renewable power sources are characterized by a power-law PSD with the Kolmogorov exponent $-5/3$, see Ref.~\onlinecite{anvari2016}. Such a characteristic PSD allows for estimating Fourier components' amplitudes through the relation $\varepsilon(\omega)\propto{S(\omega)}^{\frac{1}{2}}$. We thus define the DVI for node $i$ given a noisy perturbation with PSD $S(\omega)$ driving node $k$ as
\begin{equation}
	\mathrm{DVI}_i^{(k)}=
	\sum_{\omega\in I_{\mathrm{res}}}{S(\omega)}^{\frac{1}{2}}\left|
	\sum_{\ell=0}^{N-1}\dfrac{\imath\omega v^{[\ell]}_kv^{[\ell]}_i}{-\omega^2+\imath\alpha\omega+\lambda^{[\ell]}}\right|.
	\label{eq:DVI}
\end{equation}
Here we assume that the response to a noisy fluctuation, as the real part of the sum of the complex responses $\dot{\Theta}_i^{(k)}(\omega)$ \eqref{eq:ResponseSinusoidal} to each Fourier component of frequency $\omega$, approaches the sum of the magnitudes of $\dot{\Theta}_i^{(k)}(\omega)$ for sufficiently long time series with length $T$.

\begin{figure}[hp]
	\centering
	\includegraphics[width=\columnwidth]{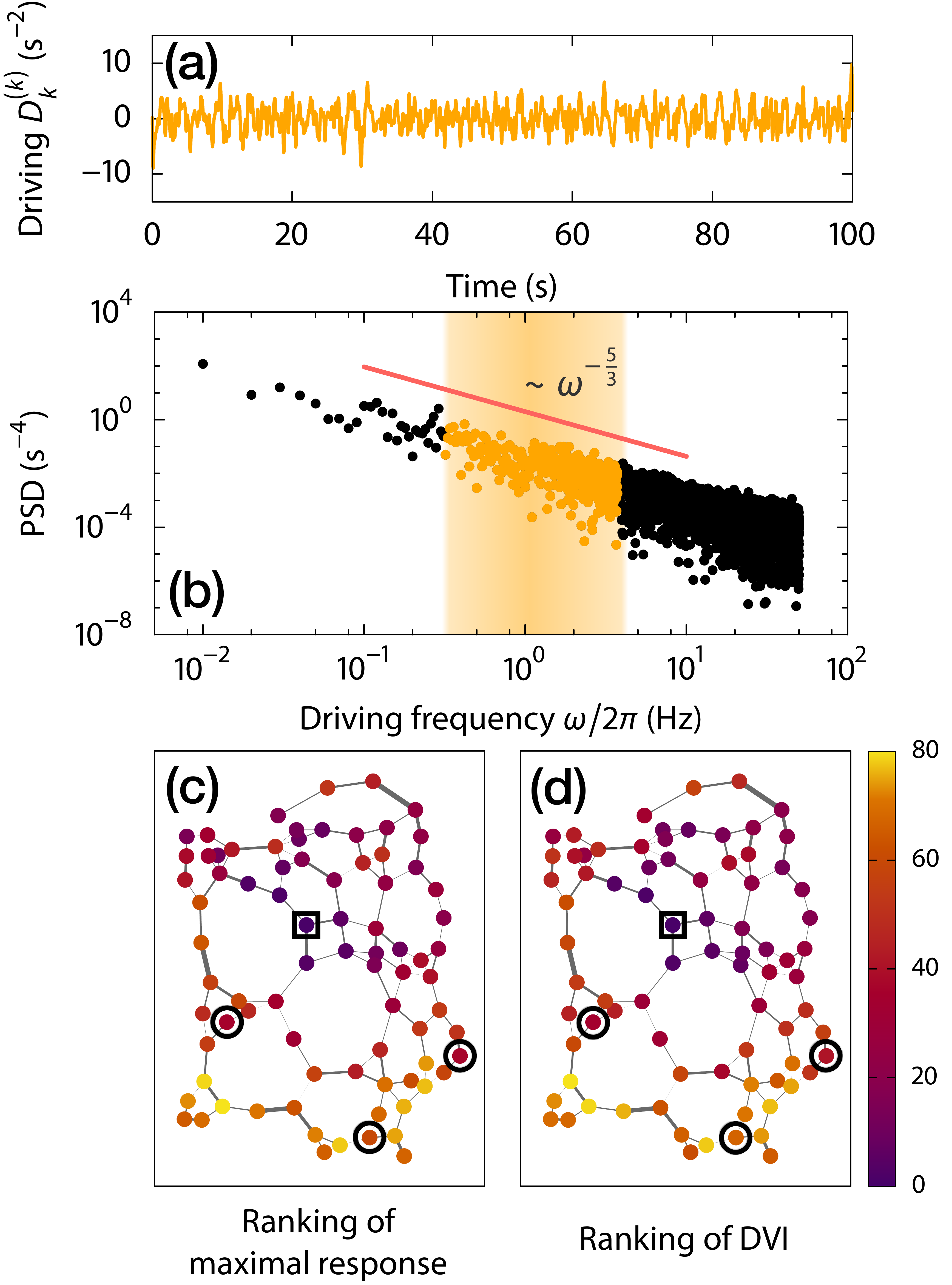}
	\caption{\textbf{The ranking of DVI well predicts the ranking of maximum resonant responses.} A time series of a resonant driving signal (a) is extracted from a colored noise with a PSD exponent $-5/3$ (b). According to the PSD of the original noise (black dots), the resonant signal is obtained by filtering the frequencies in the original signal and keeping only the ones (orange dots) in the resonance regime (shaded in orange). The ranking of the maximum frequency response $\underset{t\in [0,T]}{\max}\left|\dot\Theta_i^{(k)}(t)\right|$ in the simulation $T=100s$ is shown by a heat map in (c) and the ranking of the DVI \eqref{eq:DVI} with $S(\omega)\propto\omega^{-5/3}$ also by a heat map in (d). The perturbed node is marked with a black square. Three exemplary unexpected vulnerable nodes (marked with circles) are successfully predicted by the DVI. The network settings is the same as in Fig.~\ref{fig:resonance}. }
	\label{fig:example}
\end{figure}
\begin{figure}[h!]
	\centering
	\includegraphics[width=\columnwidth]{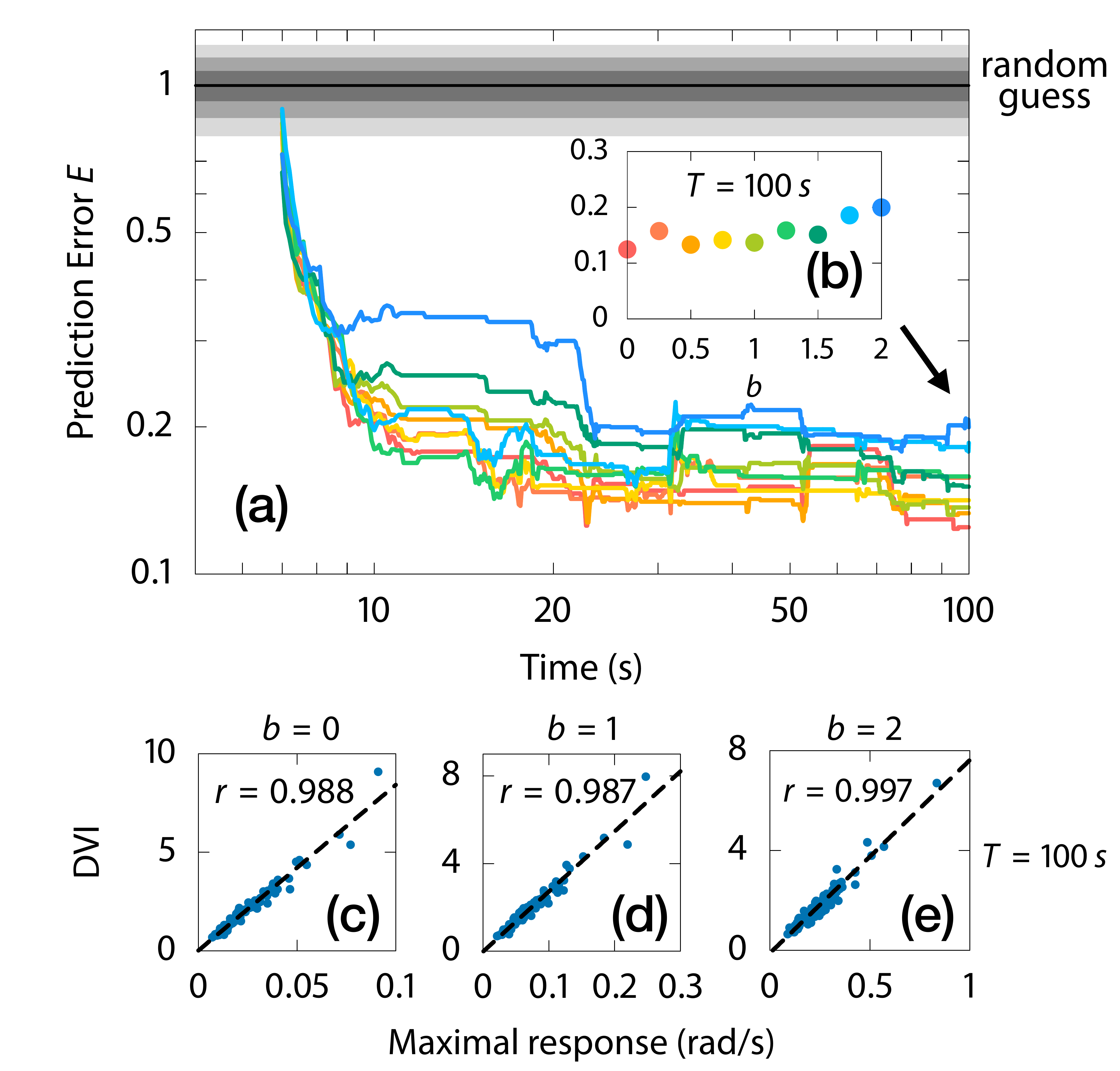}
	\caption{\textbf{Fast-converging, robustly high prediction performance of DVI ranking.} (a) The prediction error drops sharply at first, decreases over time and rests at a level about $85\%$ lower than the error of random guesses. The Gaussian error distribution of random guesses \cite{diaconis1977} is indicated by various shades of grey: black line for the expectation value of $E$ for random guesses, areas colored with different shades of grey for intervals $1\pm\rho,1\pm2\rho,1\pm3\rho$ respectively with $\rho$ being the ratio between the standard deviation and the expectation value of $E$ for random guesses. (b) A snapshot of the prediction error's dependence on signal's PSD exponent at $T=100s$. The prediction error increases slightly for larger PSD exponents of the driving signal. In both panels the prediction error is color-coded by the PSD exponent of the driving signal. (c-e) At $T=100s$, the DVI exhibits a high correlation with the maximal frequency response for various PSD exponents $b\in\{0,1,2\}$.}
	\label{fig:prediction}
\end{figure}
Note that the PSD ${S(\omega)}$ gives only the scaling information of $\varepsilon(\omega)$, so the relative value of DVI among all nodes within a network is more relevant rather than its absolute value. The ranking of DVI thus provides information about which nodes are most susceptible rather than predicting the actual response magnitudes. For instance, in a $100$-second simulation (Fig.~\ref{fig:example}), the ranking of the numerically determined maximum frequency response appears to be highly similar to the ranking given by DVI. Particularly, it gives warnings about some particular nodes (marked in circles) which are far away from the perturbation, unexpected based solely on topology (e.g. not dead-ends \cite{menck2014}), but especially vulnerable due to resonances.

We further quantitatively investigate the prediction performance of DVI in terms of its robustness over time and over the stochastic features of fluctuations. We measure DVI's prediction error with a normalized Spearman's footrule distance \cite{diaconis1977} between the predicted ranking of the maximal $\hat{\sigma}_{\mathrm{DVI}}(i)$ and the actual ranking $\sigma(i)$
\begin{equation}
	E:=\dfrac{1}{E_{\text{rand}}}\sum_{i=1}^N \left|\hat{\sigma}_{\mathrm{DVI}}(i)-\sigma(i)\right|.
	\label{eq:distance}
\end{equation}
The prediction error is normalized by the expectation value of the Spearman's footrule distance $E_{\text{rand}}=N^2/3$ between two random rankings chosen independently and uniformly in the set $S_N$ of permutation of $N$ elements, see Ref.~\onlinecite{diaconis1977}. Numerical results show that the ranking $\sigma$ of the maximum response from direct simulation converges fast to the \textit{a priori} DVI ranking $\hat\sigma_{\mathrm{DVI}}$ (Fig.~\ref{fig:prediction}a). For a $100$ second perturbation time series, we measure the true ranking $\sigma$ every $0.1$ second and compute the footrule distance $E$. For a sample power grid with the eigenfrequencies $\sim1$ Hz ($I_{\mathrm{res}}=\left[0.32\mathrm{\,Hz},3.74\mathrm{\,Hz}\right]$), the prediction error drops about $80\%$ in the first $10$ seconds and continues to decrease slowly. At $T=100\mathrm{s}$, the prediction error drops to about $15\%$ of random guess error level and the DVI is highly correlated to the maximal frequency responses with Pearson's correlation coefficient $r$ larger than $0.985$ (Fig.~\ref{fig:prediction}c-e). Furthermore, we find the prediction performance of DVI is quite robust over time and for different types of colored noise with the power-law exponent $b\in\left[0,2\right]$, from white noise to brown noise. The prediction error remains at almost the same level and shows only a mild increase with growing $b$ (Fig.~\ref{fig:prediction}b).

\section{Conclusion}
We presented a measure of dynamic node vulnerability to predict the most resonant nodes in stochastically driven oscillator networks and specifically AC power grid models. Based on a linear response theory of the network's resonance patterns for a single frequency, we propose to estimate the susceptibility of a node to stochastic driving signals by i) estimating the driving signal's Fourier spectrum by its PSD characteristics and ii) accumulating the response amplitudes to each Fourier components. Numerical results indicate strong prediction power of the proposed DVI in identifying the most resonant nodes. The true ranking of maximum response from direct simulation converges fast to the ranking prediction given by DVI, thus revealing the most vulnerable nodes. 
The prediction performance is robust not only over time, but also for various types of colored noise sources. For all tested settings, ranging from white noise to brown noise $b\in[0,2]$, the prediction error stays at a low level and the DVI ranking highly correlates with the true response ranking, with a Pearson's correlation coefficient larger than $0.985$.

Given the position and the characteristic PSD of the driving signal, for instance the location of a wind farm in a power grid network, the ranking of the DVI obtained from \eqref{eq:DVI} may help to identify which stations in the power grid would particularly be influenced by the resonant signals carried by the fluctuating wind power input, allowing precautionary measures to be taken. As the method is robust and computationally fast, it might also be applicable \textit{ad hoc} if the network changes after failures or other unforeseen events such as load shedding.

Furthermore, the DVI may support optimizing future power grid planning. For instance, new stations or new lines should be built in a way that important units in the network would not suffer from severe resonant disturbances in the altered network topology.

Taken together, the proposed dynamic vulnerability index provides a powerful tool to rank the nodal resonance level in  networks of dynamical units. Whereas the presentation above is focused on the second order Kuramoto model to overcome previous analytic limitations, the index is readily generalized, both to phase oscillator networks (with single variable nodes) and more complex systems such as networks of the third order model of power grids and more generally, networks of oscillatory and non-oscillatory dynamic units. 

\begin{acknowledgments}
We gratefully acknowledge support from the Deutsche Forschungsgemeinschaft (DFG; German Research Foundation) under Germany’s Excellence Strategy EXC-2068 390729961 Cluster of Excellence Physics of Life and the Cluster of Excellence Center for Advancing Electronics at TU Dresden, the German Federal Ministry for Research and Education (BMBF grants no. 03SF0472F and 03EK3055F), and the Helmholtz Association (via the joint initiative “Energy System 2050—A Contribution of the Research Field Energy” and grant no. VH-NG-1025). 
\end{acknowledgments}

\bibliography{dvi}
\end{document}